\documentclass[RNAAS]{aastex62}
\usepackage{graphicx,epsf,soul,xcolor,gensymb, mathtools}
\usepackage{rotating}
\usepackage{multirow}

\def\sn{AT\,2022tsd}

\begin{document}

\title{Unprecedented X-ray Emission from the Fast Blue Optical Transient \sn\,}

\correspondingauthor{David Matthews}
\email{davidjacobmatthews@gmail.com}

\author[0000-0002-4513-3849]{D. J. Matthews}
\affiliation{Department of Astronomy, University of California, Berkeley, CA 94720, USA}

\author[0000-0003-4768-7586]{R.~Margutti}
\affiliation{Department of Astronomy, University of California, Berkeley, CA 94720, USA}
\affiliation{Department of Physics, University of California, Berkeley, CA 94720, USA}

\author[0000-0002-4670-7509]{B. D.~Metzger}
\affil{Department of Physics and Columbia Astrophysics Laboratory, Columbia University, New York, NY 10027, USA}
\affil{Center for Computational Astrophysics, Flatiron Institute, 162 5th Ave, New York, NY 10010, USA}

\author[0000-0002-0763-3885]{D. Milisavljevic}
\affiliation{Purdue University, Department of Physics and Astronomy, 525 Northwestern Ave, West Lafayette, IN 47907 }
\affiliation{Integrative Data Science Initiative, Purdue University, West Lafayette, IN 47907, USA}

\author[0000-0003-0216-8053]{G. Migliori}
\affiliation{Dipartamento di Fisica e Astronomia, Universita di Bologna, viale Berti Pichat 6/2, I-40127 Bologna, Italy}
\affiliation{INAF – Istituto di Radioastronomia, Via P. Gobetti 101, I-40129 Bologna, Italy}

\author[0000-0003-1792-2338]{T. Laskar}
\affiliation{University of Utah, Department of Physics \& Astronomy, 115 S 1400 E, Salt Lake City, UT 84112, USA}

\author[0000-0001-6415-0903]{D. Brethauer}
\affiliation{Department of Astronomy, University of California, Berkeley, CA 94720, USA}

\author[0000-0002-9392-9681]{E. Berger}
\affiliation{Center for Astrophysics, Harvard \& Smithsonian, 60 Garden Street, Cambridge, MA 02138-1516, USA}

\author[0000-0002-7706-5668]{R.~Chornock}
\affiliation{Department of Astronomy, University of California, Berkeley, CA 94720, USA}

\author[0000-0001-7081-0082]{M. Drout}
\affiliation{David A. Dunlap Department of Astronomy \& Astrophysics, University of Toronto, 50 St. George Street, Toronto, ON M5S 3H4, Canada}
\affiliation{Dunlap Institute for Astronomy \& Astrophysics, University of Toronto, 50 St. George Street, Toronto, ON M5S 3H4, Canada}

\author[0000-0003-2558-3102]{E. Ramirez-Ruiz}
\affiliation{Niels Bohr Institute, University of Copenhagen, Blegdamsvej 17, DK-2100 Copenhagen, Denmark}
\affiliation{Department of Astronomy \& Astrophysics, University of California, Santa Cruz, CA 95064, USA}

\begin{abstract}
We present the X-ray monitoring campaign of \sn\, in the time range $\delta t_{rest} = 23 - 116$\,d rest-frame since discovery. With an initial 0.3-10 keV X-ray luminosity of $L_x \approx 10^{44}$ erg s$^{-1}$ at $\delta t_{rest}\approx$\,23\,d, \sn\, is the most luminous FBOT to date and rivals even the most luminous GRBs. We find no statistical evidence for spectral evolution. The average X-ray spectrum is well described by an absorbed simple power-law spectral model with best-fitting photon index $\Gamma = 1.89 ^{+0.09}_{-0.08}$ and marginal evidence at the 3$\sigma$ confidence level for intrinsic absorption $NH_{int}\approx 4\times10^{19}$ cm$^{-2}$. The X-ray light-curve behavior can be either interpreted as a power-law decay $L_x\propto t^{\alpha}$ with $\alpha\approx -2$ and superimposed X-ray variability, or as a broken power-law with a steeper post-break decay as observed in other FBOTs such as AT\,2018cow. We briefly compare these results to accretion models of TDEs and GRB afterglow models.


\end{abstract}

\section{Introduction}
Fast Blue Optical Transients (FBOTs) are an enigmatic class of explosive transients identified by optical transient surveys over the past decade. FBOTs are characterized by an extremely rapid rise to maximum light ($t_{\rm rise}<10$ days), luminous emission ($L_{\rm p}>10^{43}\,\rm{erg\,s^{-1}}$) and blue colors (e.g., \citealt{drout2014, tanaka2016, Tampo20,ho2021}). Traditional supernova (SN) models cannot easily explain the short timescales, high peak luminosities, and lack of UV line blanketing observed in many of these transients.

The FBOT \sn\, (aka ZTF22abftjko) was discovered on September 7th 2022 at 11:21 UT by the Zwicky Transient Facility (ZTF) and reported by the Automatic Learning for the Rapid Classification of Events (ALeRCE, \citealt{discovery}). The discovery was corroborated by a later report from the Gravitational-Wave Optical Transient Observer (GOTO) collaboration with their own earlier detection on September 6th 2022 at 05:25 UT \citep{astronote208}. The field of \sn\, was also observed
with the Panoramic Survey Telescope and Rapid Response System (Pan-STARRS), and forced photometry revealed no historical activity at the position of the transient \citep{astronote206}.  We thus adopt September 6th 2022 (MJD 59828) as our $T
_0$. Follow-up observations with the Keck Low Resolution Imaging Spectrometer (LRIS) on September 23rd 2022 revealed a blue spectral continuum with narrow emission lines, consistent with a host galaxy at a redshift of $z=0.256$ \citep{astronote199}. This placed the discovery absolute magnitude at $M=-20.0$ at a wavelength close to the rest-frame g-band, comparable with the population of optically luminous FBOTs (LFBOTs hereafter). LFBOTs are known to be associated with luminous X-ray and radio emission \citep{coppejans20, ho2021}. 

On October 2nd 2022 ($\delta t_{rest}\approx25$\,d), the Very Large Array (VLA) detected radio emission with $F_{\nu}=$25 $\mu$Jy at 15 GHz, corresponding to  $L_{\nu}=5\times10^{28}$ erg s$^{-1}$ Hz$^{-1}$  \citep{astronote205}, which is 
comparable to that of AT\,2018cow ($3\times10^{28}$ erg s$^{-1}$ Hz$^{-1}$  \citep{margutti2019}
and lower than that of AT\,2022xnd ($L_{\nu}=2\times10^{29}$ erg s$^{-1}$ Hz$^{-1}$) at a similar epoch \citep{ho2022}. 
\sn\, was first detected in the X-rays on October 4th 2022 ($
\delta t_{rest}=$\,22.8\,d) by the Neil Gehrels Swift Observatory X-ray Telescope (Swift-XRT) with an inferred 0.3-10 keV luminosity of 
$L_x=10^{44}$ erg s$^{-1}$ \citep{astronote207}. An additional optical follow-up imaging sequence on December 15th 2022 ($
\delta t_{rest}=$\,80\,d)  revealed a minute-timescale optical flare at the location of the transient  which is unprecedented among FBOTs \citep{astronote267}.  In this research note, we present the results of the follow-up X-ray campaign of \sn\, with \emph{Swift}-XRT and the Chandra X-ray Observatory (CXO).

Times are reported with respect of $T_0$ and in the rest frame, unless explicitly noted.  We adopt a luminosity distance of $d_L = 1.3 $ Gpc, assuming standard $\Lambda$CDM cosmology ($H_{0}$ = 70 km s$^{-1}$ Mpc$^{-1}$, $\Omega_M = 0.27$, $\Omega_{\Lambda} = 0.73$). 

\section{X-ray data analysis}\label{Sec:data}
\subsection{Swift-XRT} \label{SubSec:XRT}
\emph{Swift}-XRT observed \sn\, from October 4th  to December 16th 2022 ($\delta t_{rest}=$ 23-80\,d) for a total exposure of 39.6\,ks (PI Schulze). We reduced the data following standard practice (e.g, \citealt{Margutti13}). An X-ray source is clearly detected at the location of \sn\, with initial count-rate of $\approx 0.01\rm{s^{-1}}$ (0.3-10\,keV). The spectrum is well fit by an absorbed power-law model with photon index $\Gamma = 2.00^{+0.17}_{-0.15}$ and no evidence for intrinsic absorption. The Galactic neutral hydrogen column density in the direction of the transient is   $NH_g=2.1\times10^{21}\,\rm{cm^{-2}}$ \citep{willingale13}, and we use updated solar abundances from \citet{aspl09}. We account for the low-count statistics by using the {\tt cstat} statistic function in the software package {\tt Xspec} \citep{arnaud1996} and we infer uncertainties with Markov chain Monte Carlo (MCMC) simulations. Reported uncertainties are $1\,\sigma$ Gaussian equivalent. 

\begin{figure*}[t!]
    \centering    
    \raisebox{0.15cm}{\includegraphics[width=0.48\linewidth]{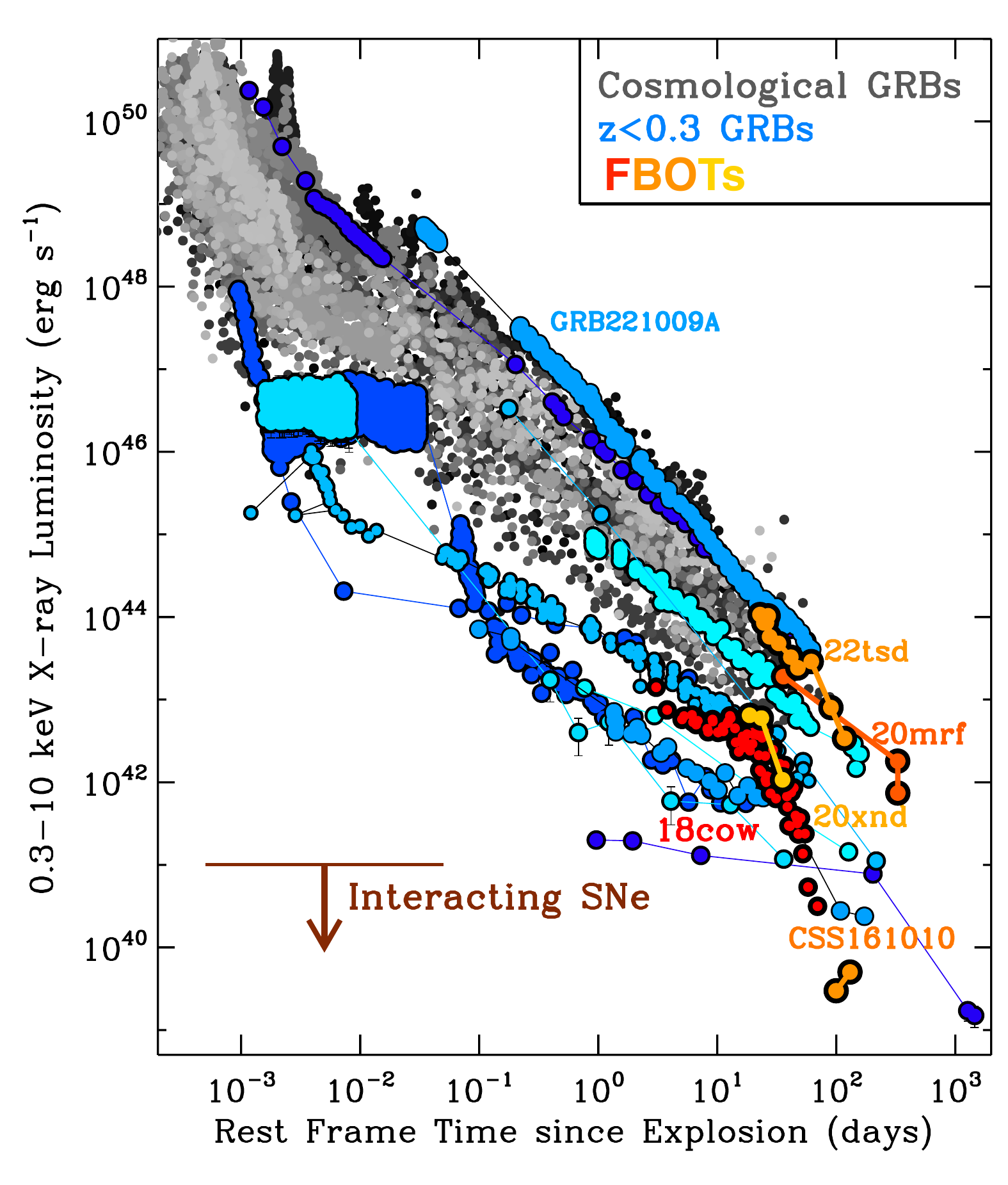}}
    \includegraphics[width=0.48\textwidth]{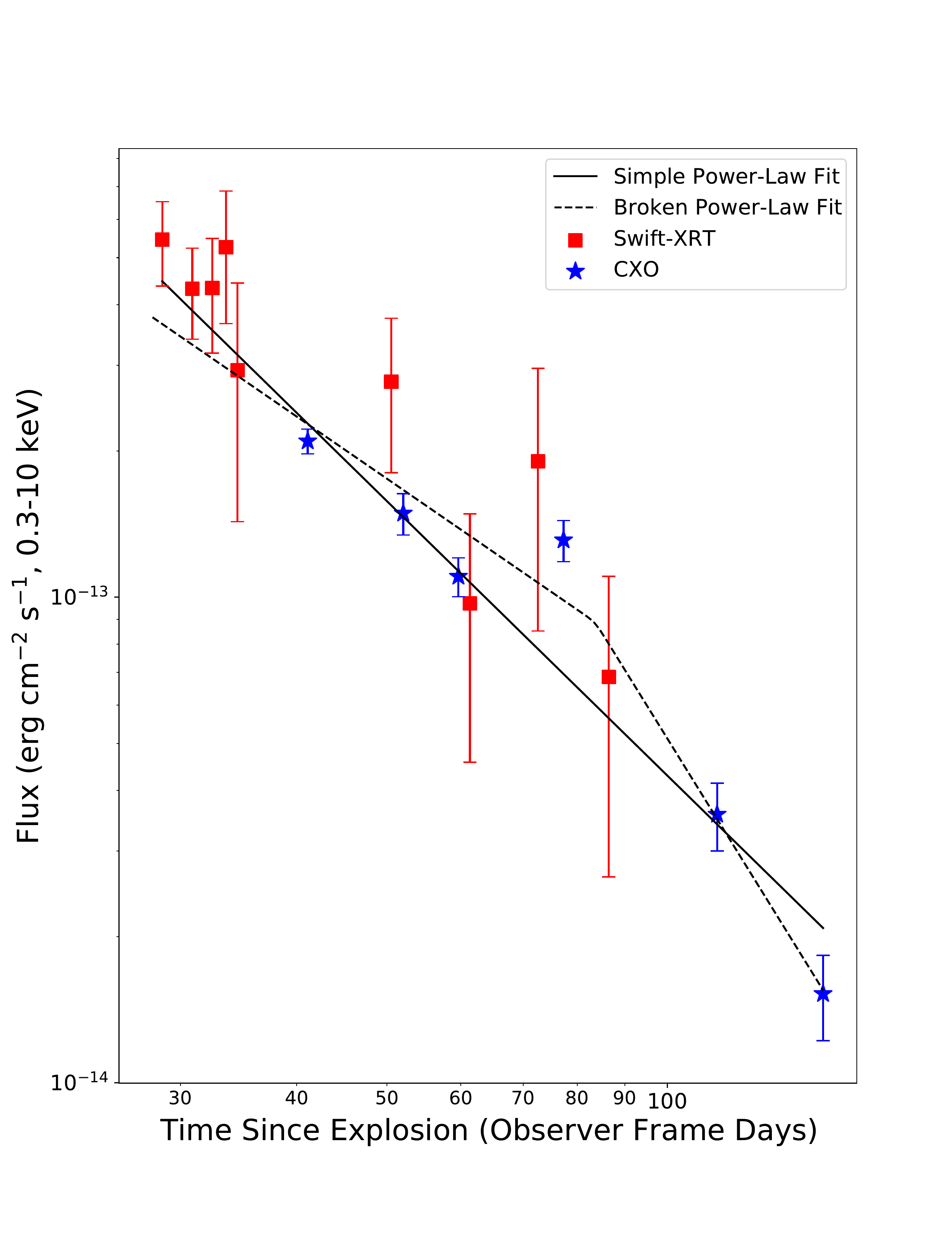}
    \caption{Left: X-rays from \sn\, in the context of other FBOTs and comparable GRBs. Right: X-ray light curve of \sn\,. The solid line shows the best-fit simple power-law with $\alpha=-1.88\pm15$. The dashed line shows the best-fit broken power-law, with $t_{\rm{break}} = 10^{(6.86\pm0.15)}$ seconds (corresponding to $\sim 83.7$ days), $\alpha_1 = -1.32\pm0.23$, and $\alpha_2=-3.10 \pm1.45$. We note that the fourth CXO epoch is an outlier in both models, possibly indicating central engine variability, and emphasizing the importance of high-cadence X-ray monitoring.}
    \label{Fig:Xray_LC}
\end{figure*}

\subsection{Chandra X-ray Observatory (CXO)} \label{SubSec:CXO}

We acquired five epochs of CXO observations under Chandra program \#24500280 (PI Matthews), and an additional epoch was obtained through a Chandra DDT (PI Ho, program \#503444). These observations began on October 16th 2022 and the most recent epoch was taken on January 30th 2023 ($\delta t_{rest}=$\,32-116\,d) for a total exposure of 153 ks. We used the software package {\tt CIAO} (v4.13) to reprocess and reduce the data, applying standard ACIS data filtering between 0.5-8 keV \citep{Fruscione06}. We found evidence for statistically significant X-ray emission consistent with the location of \sn\, after re-aligning the images to a common astrometric solution. For each observation, we used {\tt specextract} to extract a spectrum with a $1.5\arcsec$ source region and $35\arcsec$ source-free background region, and modeled each spectrum with an absorbed power-law model in {\tt Xspec}. We employed {\tt cstat} and we used MCMC simulations to properly estimate the uncertainties.

The analysis of the individual CXO epochs and a comparison with the early \emph{Swift}-XRT data showed no significant spectral evolution. Similarly, we investigated the spectral properties of \sn\, before and after the possible temporal break in the light curve (see Fig. \ref{Fig:Xray_LC}) and found no statistical evidence for spectral evolution of the source. We thus proceeded by jointly fitting the \emph{CXO} and \emph{Swift}-XRT data with an absorbed simple power-law model allowing for different normalizations. From this joint fit we found marginal evidence (i.e. at the $3\sigma$ confidence level) for intrinsic absorption $NH_{int}\approx 4\times 10^{19}\,\rm{cm^{-2}}$. The best-fitting photon index is $\Gamma=1.89^{+0.09}_{-0.08}$. We adopt these values for the flux calibration of the entire data set (see Table 1). 


\section{Discussion and conclusions}\label{Sec:discconc}
With $L_x \approx 10^{44}$ erg s$^{-1}$ when first observed at $\delta t_{rest}\approx 23$\,d, \sn\, is the most luminous FBOTs in the X-rays discovered so far. Being more luminous than most Gamma-Ray Bursts (GRBs) at similar times, \sn\, rivals in luminosity the record-breaking GRB\,221009A (\citealt{williams23}, Fig. \ref{Fig:Xray_LC}, left panel). We fit the X-ray light curve of \sn\, with both a simple power-law model as well as a smoothed broken power-law (as in \citealt{demarchi23}), where we freeze the smoothing parameter as $s=-0.01$ for a sharp break since the break smoothing is not constrained. We find the best-fit parameters for the simple power-law model to be a slope of $\alpha = -1.88 \pm 0.15$ and a normalization factor of $F=1.12 \pm 0.07 \times 10^{-13}$ erg cm$^{-2}$ s$^{-1}$ (normalized at $t=60$ days). For the smoothed broken power-law model, we find the best-fit parameters to be $F_{\rm{break}} = 10^{(-13.05\pm0.22)}$ erg cm$^{-2}$ s$^{-1}$, $t_{\rm{break}} = 10^{(6.86\pm0.15)}$ seconds (corresponding to $\sim 83.7$ days), $\alpha_1 = -1.32\pm0.23$, and $\alpha_2=-3.10 \pm1.45$.

To compare these models, we run an F-test with the null hypothesis that our data are more likely to originate from a simple power-law model. Using the $\chi^2$ statistic for each model we find a p-value of $p=0.46$ for the F-test, indicating that the smoothed broken power-law model does not represent a statistically significant improvement over the simple power-law model, despite the former having a better $\chi^2$ statistic. However we do note that the pre-break and post-break power-law slopes of the smoothed broken power-law model are comparable to those of AT 2018cow and AT 2020xnd, which were both consistent with scalings of $t^{-1}$ and $t^{-4}$ \citep{margutti2019,bright22}. The pre-break power-law slope of AT 2020mrf ($t^{-1.3}$) is consistent with our results, however there is no available comparison for the post-break slope of AT 2020mrf \citep{yao22}. Our results are also consistent with tidal disruption event (TDE) fallback accretion models where the decay rate scales as $t^{-5/3}$ \citep{rees88,perley19}, as well as with Gamma-Ray Burst (GRB) post jet-break models where the decay rate scales as $t^{-p}$ after the jet-break where $p=2$ \citep{wang18}.  Alternatively, the accretion rate onto a compact object from a viscously spreading thick accretion disk, such as that created by rotating stellar core collapse (e.g., \citealt{Gottlieb+22}) or a helium star/compact object merger (e.g., \citealt{Fryer&Woosley98,Metzger22}), is predicted to decay as $t^{-4/3}$ or steeper (e.g., \citealt{Metzger+08}).  We conclude that, while some of the observed X-ray properties of \sn\, are comparable to those of other FBOTs, its unprecedented X-ray luminosity establishes an extreme case among an already extreme class of explosive transients.

\bibliography{ref}

\begin{thebibliography}{}
\expandafter\ifx\csname natexlab\endcsname\relax\def\natexlab#1{#1}\fi

\bibitem[{{Arnaud}(1996)}]{arnaud1996}
{Arnaud}, K.~A. 1996, in Astronomical Society of the Pacific Conference Series,
  Vol. 101, Astronomical Data Analysis Software and Systems V, ed. G.~H.
  {Jacoby} \& J.~{Barnes}, 17

\bibitem[{{Asplund} {et~al.}(2009){Asplund}, {Grevesse}, {Sauval}, \&
  {Scott}}]{aspl09}
{Asplund}, M., {Grevesse}, N., {Sauval}, A.~J., \& {Scott}, P. 2009, \araa, 47,
  481

\bibitem[{Bright {et~al.}(2022)Bright, Margutti, Matthews, Brethauer,
  Coppejans, Wieringa, Metzger, DeMarchi, Laskar, Romero, Alexander, Horesh,
  Migliori, Chornock, Berger, Bietenholz, Devlin, Dicker, Jacobson-Gal{\'{a}
  }n, Mason, Milisavljevic, Motta, Mroczkowski, Ramirez-Ruiz, Rhodes, Sarazin,
  Sfaradi, \& Sievers}]{bright22}
Bright, J.~S., Margutti, R., Matthews, D., {et~al.} 2022, The Astrophysical
  Journal, 926, 112

\bibitem[{{Coppejans} {et~al.}(2022){Coppejans}, {Lyman}, {Bright}, {Ulaczyk},
  {Ackley}, {Dyer}, {Steeghs}, {Galloway}, {Dhillon}, {O'Brien}, {Ramsay},
  {Pollacco}, {Thrane}, {Noysena}, {Kotak}, {Nuttall}, {Pall{\'e}}, {Breton},
  {Killestein}, {O'Neill}, {Kumar}, \& {Jiminez-Ibarra}}]{astronote208}
{Coppejans}, D., {Lyman}, J., {Bright}, J., {et~al.} 2022, Transient Name
  Server AstroNote, 208, 1

\bibitem[{Coppejans {et~al.}(2020)Coppejans, Margutti, Terreran, Nayana,
  Coughlin, Laskar, Alexander, Bietenholz, Caprioli, Chandra, Drout, Frederiks,
  Frohmaier, Hurley, Kochanek, MacLeod, Meisner, Nugent, Ridnaia, Sand,
  Svinkin, Ward, Yang, Baldeschi, Chilingarian, Dong, Esquivia, Fong, Guidorzi,
  Lundqvist, Milisavljevic, Paterson, Reichart, Shappee, Stroh, Valenti,
  Zauderer, \& Zhang}]{coppejans20}
Coppejans, D.~L., Margutti, R., Terreran, G., {et~al.} 2020, The Astrophysical
  Journal Letters, 895, L23

\bibitem[{DeMarchi {et~al.}(2023)DeMarchi, Finstad, \& Margutti}]{demarchi23}
DeMarchi, L., Finstad, D., \& Margutti, R. 2023, Research Notes of the AAS, 7,
  77

\bibitem[{{Drout} {et~al.}(2014){Drout}, {Chornock}, {Soderberg}, {Sanders},
  {McKinnon}, {Rest}, {Foley}, {Milisavljevic}, {Margutti}, {Berger},
  {Calkins}, {Fong}, {Gezari}, {Huber}, {Kankare}, {Kirshner}, {Leibler},
  {Lunnan}, {Mattila}, {Marion}, {Narayan}, {Riess}, {Roth}, {Scolnic},
  {Smartt}, {Tonry}, {Burgett}, {Chambers}, {Hodapp}, {Jedicke}, {Kaiser},
  {Magnier}, {Metcalfe}, {Morgan}, {Price}, \& {Waters}}]{drout2014}
{Drout}, M.~R., {Chornock}, R., {Soderberg}, A.~M., {et~al.} 2014, ApJ, 794, 23

\bibitem[{{Fruscione} {et~al.}(2006){Fruscione}, {McDowell}, {Allen},
  {Brickhouse}, {Burke}, {Davis}, {Durham}, {Elvis}, {Galle}, {Harris},
  {Huenemoerder}, {Houck}, {Ishibashi}, {Karovska}, {Nicastro}, {Noble},
  {Nowak}, {Primini}, {Siemiginowska}, {Smith}, \& {Wise}}]{Fruscione06}
{Fruscione}, A., {McDowell}, J.~C., {Allen}, G.~E., {et~al.} 2006, in SPIE,
  Vol. 6270, SPIE, ed. D.~R. {Silva} \& R.~E. {Doxsey}, 62701V

\bibitem[{{Fryer} \& {Woosley}(1998)}]{Fryer&Woosley98}
{Fryer}, C.~L., \& {Woosley}, S.~E. 1998, \apjl, 502, L9

\bibitem[{{Fulton} {et~al.}(2022){Fulton}, {Smartt}, {Smith}, {Young},
  {Srivastav}, {Moore}, {Chambers}, {Huber}, {Schultz}, {Boer}, {Bulger},
  {Fairlamb}, {Gao}, {Lin}, {Lowe}, {Magnier}, {Wainscoat}, {Chen}, {Rest}, \&
  {Stubbs}}]{astronote206}
{Fulton}, M., {Smartt}, S.~J., {Smith}, K.~W., {et~al.} 2022, Transient Name
  Server AstroNote, 206, 1

\bibitem[{{Gottlieb} {et~al.}(2022){Gottlieb}, {Tchekhovskoy}, \&
  {Margutti}}]{Gottlieb+22}
{Gottlieb}, O., {Tchekhovskoy}, A., \& {Margutti}, R. 2022, \mnras, 513, 3810

\bibitem[{{Ho} \& {Perley}(2022)}]{astronote205}
{Ho}, A.~Y.~Q., \& {Perley}, D.~A. 2022, Transient Name Server AstroNote, 205,
  1

\bibitem[{{Ho} {et~al.}(2022{\natexlab{a}}){Ho}, {Perley}, {Chen}, {Schulze},
  {Sollerman}, \& {Gal-Yam}}]{astronote267}
{Ho}, A.~Y.~Q., {Perley}, D.~A., {Chen}, P., {et~al.} 2022{\natexlab{a}},
  Transient Name Server AstroNote, 267, 1

\bibitem[{{Ho} {et~al.}(2022{\natexlab{b}}){Ho}, {Perley}, {Filippenko},
  {Zheng}, {Brink}, {Li}, \& {Wang}}]{astronote199}
{Ho}, A.~Y.~Q., {Perley}, D.~A., {Filippenko}, A.~V., {et~al.}
  2022{\natexlab{b}}, Transient Name Server AstroNote, 199, 1

\bibitem[{{Ho} {et~al.}(2021){Ho}, {Perley}, {Gal-Yam}, {Lunnan}, {Sollerman},
  {Schulze}, {Das}, {Dobie}, {Yao}, {Fremling}, {Adams}, {Anand}, {Andreoni},
  {Bellm}, {Bruch}, {Burdge}, {Castro-Tirado}, {Dahiwale}, {De}, {Dekany},
  {Drake}, {Duev}, {Graham}, {Helou}, {Kaplan}, {Karambelkar}, {Kasliwal},
  {Kool}, {Kulkarni}, {Mahabal}, {Medford}, {Miller}, {Nordin}, {Ofek},
  {Petitpas}, {Riddle}, {Sharma}, {Smith}, {Stewart}, {Taggart}, {Tartaglia},
  {Tzanidakis}, \& {Winters}}]{ho2021}
{Ho}, A. Y.~Q., {Perley}, D.~A., {Gal-Yam}, A., {et~al.} 2021, arXiv e-prints,
  arXiv:2105.08811

\bibitem[{{Ho} {et~al.}(2022{\natexlab{c}}){Ho}, {Margalit}, {Bremer},
  {Perley}, {Yao}, {Dobie}, {Kaplan}, {O'Brien}, {Petitpas}, \& {Zic}}]{ho2022}
{Ho}, A. Y.~Q., {Margalit}, B., {Bremer}, M., {et~al.} 2022{\natexlab{c}},
  \apj, 932, 116

\bibitem[{{Margutti} {et~al.}(2013){Margutti}, {Zaninoni}, {Bernardini},
  {Chincarini}, {Pasotti}, {Guidorzi}, {Angelini}, {Burrows}, {Capalbi},
  {Evans}, {Gehrels}, {Kennea}, {Mangano}, {Moretti}, {Nousek}, {Osborne},
  {Page}, {Perri}, {Racusin}, {Romano}, {Sbarufatti}, {Stafford}, \&
  {Stamatikos}}]{Margutti13}
{Margutti}, R., {Zaninoni}, E., {Bernardini}, M.~G., {et~al.} 2013, \mnras,
  428, 729

\bibitem[{{Margutti} {et~al.}(2019){Margutti}, {Metzger}, {Chornock}, {Vurm},
  {Roth}, {Grefenstette}, {Savchenko}, {Cartier}, {Steiner}, {Terreran},
  {Margalit}, {Migliori}, {Milisavljevic}, {Alexander}, {Bietenholz},
  {Blanchard}, {Bozzo}, {Brethauer}, {Chilingarian}, {Coppejans}, {Ducci},
  {Ferrigno}, {Fong}, {G{\"o}tz}, {Guidorzi}, {Hajela}, {Hurley}, {Kuulkers},
  {Laurent}, {Mereghetti}, {Nicholl}, {Patnaude}, {Ubertini}, {Banovetz},
  {Bartel}, {Berger}, {Coughlin}, {Eftekhari}, {Frederiks}, {Kozlova},
  {Laskar}, {Svinkin}, {Drout}, {MacFadyen}, \& {Paterson}}]{margutti2019}
{Margutti}, R., {Metzger}, B.~D., {Chornock}, R., {et~al.} 2019, \apj, 872, 18

\bibitem[{{Metzger}(2022)}]{Metzger22}
{Metzger}, B.~D. 2022, \apj, 932, 84

\bibitem[{{Metzger} {et~al.}(2008){Metzger}, {Piro}, \&
  {Quataert}}]{Metzger+08}
{Metzger}, B.~D., {Piro}, A.~L., \& {Quataert}, E. 2008, \mnras, 390, 781

\bibitem[{{Munoz-Arancibia} {et~al.}(2022){Munoz-Arancibia}, {Bauer},
  {Forster}, {Pignata}, {Mourao}, {Hernandez-Garcia}, {Galbany},
  {Silva-Farfan}, {Hoshino}, {Camacho}, {Arredondo}, {Cabrera-Vives},
  {Carrasco-Davis}, {Estevez}, {Huijse}, {Reyes}, {Reyes}, {Sanchez-Saez},
  {Rodriguez-Mancini}, {Catelan}, {Eyheramendy}, \& {Graham}}]{discovery}
{Munoz-Arancibia}, A., {Bauer}, F.~E., {Forster}, F., {et~al.} 2022, Transient
  Name Server Discovery Report, 2022-2602, 1

\bibitem[{{Perley} {et~al.}(2019){Perley}, {Mazzali}, {Yan}, {Cenko}, {Gezari},
  {Taggart}, {Blagorodnova}, {Fremling}, {Mockler}, {Singh}, {Tominaga},
  {Tanaka}, {Watson}, {Ahumada}, {Anupama}, {Ashall}, {Becerra}, {Bersier},
  {Bhalerao}, {Bloom}, {Butler}, {Copperwheat}, {Coughlin}, {De}, {Drake},
  {Duev}, {Frederick}, {Gonz{\'a}lez}, {Goobar}, {Heida}, {Ho}, {Horst},
  {Hung}, {Itoh}, {Jencson}, {Kasliwal}, {Kawai}, {Khanam}, {Kulkarni},
  {Kumar}, {Kumar}, {Kutyrev}, {Lee}, {Maeda}, {Mahabal}, {Murata}, {Neill},
  {Ngeow}, {Penprase}, {Pian}, {Quimby}, {Ramirez-Ruiz}, {Richer},
  {Rom{\'a}n-Z{\'u}{\~n}iga}, {Sahu}, {Srivastav}, {Socia}, {Sollerman},
  {Tachibana}, {Taddia}, {Tinyanont}, {Troja}, {Ward}, {Wee}, \&
  {Yu}}]{perley19}
{Perley}, D.~A., {Mazzali}, P.~A., {Yan}, L., {et~al.} 2019, \mnras, 484, 1031

\bibitem[{{Rees}(1988)}]{rees88}
{Rees}, M.~J. 1988, \nat, 333, 523

\bibitem[{{Schulze} {et~al.}(2022){Schulze}, {Ho}, {Perley}, {Yan}, \&
  {Fremling}}]{astronote207}
{Schulze}, S., {Ho}, A.~Y.~Q., {Perley}, D.~A., {Yan}, L., \& {Fremling}, C.
  2022, Transient Name Server AstroNote, 207, 1

\bibitem[{{Tampo} {et~al.}(2020){Tampo}, {Tanaka}, {Maeda}, {Yasuda},
  {Tominaga}, {Jiang}, {Moriya}, {Morokuma}, {Suzuki}, {Takahashi}, {Kokubo},
  \& {Kawana}}]{Tampo20}
{Tampo}, Y., {Tanaka}, M., {Maeda}, K., {et~al.} 2020, \apj, 894, 27

\bibitem[{{Tanaka} {et~al.}(2016){Tanaka}, {Tominaga}, {Morokuma}, {Yasuda},
  {Furusawa}, {Baklanov}, {Blinnikov}, {Moriya}, {Doi}, {Jiang}, {Kato},
  {Kikuchi}, {Kuncarayakti}, {Nagao}, {Nomoto}, \& {Taniguchi}}]{tanaka2016}
{Tanaka}, M., {Tominaga}, N., {Morokuma}, T., {et~al.} 2016, ApJ, 819, 5

\bibitem[{Wang {et~al.}(2018)Wang, Zhang, Liang, Lu, Lin, Li, \& Li}]{wang18}
Wang, X.-G., Zhang, B., Liang, E.-W., {et~al.} 2018, The Astrophysical Journal,
  859, 160

\bibitem[{Williams {et~al.}(2023)Williams, Kennea, Dichiara, Kobayashi,
  Iwakiri, Beardmore, Evans, Heinz, Lien, Oates, Negoro, Cenko, Buisson,
  Hartmann, Jaisawal, Kuin, Lesage, Page, Parsotan, Pasham, Sbarufatti, Siegel,
  Sugita, Younes, Ambrosi, Arzoumanian, Bernardini, Campana, Capalbi, Caputo,
  D'A{\`{\i} }, D'Avanzo, D'Elia, Pasquale, Eyles-Ferris, Ferrara, Gendreau,
  Gropp, Kawai, Klingler, Laha, Melandri, Mihara, Moss, O'Brien, Osborne,
  Palmer, Perri, Serino, Sonbas, Stamatikos, Starling, Tagliaferri, Tohuvavohu,
  Zane, \& Ziaeepour}]{williams23}
Williams, M.~A., Kennea, J.~A., Dichiara, S., {et~al.} 2023, The Astrophysical
  Journal Letters, 946, L24

\bibitem[{{Willingale} {et~al.}(2013){Willingale}, {Starling}, {Beardmore},
  {Tanvir}, \& {O'Brien}}]{willingale13}
{Willingale}, R., {Starling}, R.~L.~C., {Beardmore}, A.~P., {Tanvir}, N.~R., \&
  {O'Brien}, P.~T. 2013, \mnras, 431, 394

\bibitem[{Yao {et~al.}(2022)Yao, Ho, Medvedev, J., Perley, Kulkarni, Chandra,
  Sazonov, Gilfanov, Khorunzhev, Khatami, \& Sunyaev}]{yao22}
Yao, Y., Ho, A. Y.~Q., Medvedev, P., {et~al.} 2022, The Astrophysical Journal,
  934, 104

\end{thebibliography}
\section{Appendix}
\label{SubSec:Appendix}
\begin{deluxetable}{ccccccc}[h]
\tablecaption{X-ray Observations of \sn. All fluxes are given in the 0.3-10 keV energy range with 1$\sigma$ error bars.}
\label{Tab:xray_data}
\tablehead{
\colhead{MJD Obs} & \colhead{T$_{\rm{rest}}$} & \colhead{Exposure} & \colhead{Absorbed Flux} & \colhead{Unabsorbed Flux} & \multirow{2}{*}{Telescope/Instrument} & \multirow{2}{*}{PI} \\
(UT) & (Days) & (ks) & ($10^{-13}\rm{erg\,s^{-1}cm^{-2}}$) & ($10^{-13}\rm{erg\,s^{-1}cm^{-2}}$) & &  }
\startdata
59856.7	&	22.8	&	3.7	&		$5.24^{+1.04}_{-1.04}$	&	$7.47	^{+	1.47	}_{	-1.47	}$	&	Swift/XRT	&	Schulze$^1$ \\
59858.9	&	24.6	&	3.8	&		$4.15^{+0.88}_{-0.88}$	&	$5.92	^{+	1.26	}_{	-1.26	}$	&	Swift/XRT	&	Schulze$^1$ \\
59860.5	&	25.8	&	2.5	&		$4.17^{+1.11}_{-1.11}$	&	$5.94	^{+	1.57	}_{	-1.57	}$	&	Swift/XRT	&	Schulze$^1$ \\
59861.6	&	26.7	&	2.3	&		$5.06^{+1.54}_{-1.29}$	&	$7.22	^{+	2.20	}_{	-1.84	}$	&	Swift/XRT	&	Schulze$^1$ \\
59862.5	&	27.5	&	2.4	&		$2.82^{+1.45}_{-1.10}$	&	$4.03	^{+	2.06	}_{	-1.57	}$	&	Swift/XRT	&	Schulze$^1$ \\
59869.1	&	32.7	&	19.3	&		$2.35^{+0.16}_{-0.15}$	&	$3.81	^{+	1.34	}_{	-1.09	}$	&	CXO/ACIS-S	&	Matthews$^2$ \\
59878.2	&	40	&	2.8	&		$2.78^{+0.97}_{-0.79}$	&	$1.33	^{+	0.71	}_{	-0.54	}$	&	Swift/XRT	&	Schulze$^1$ \\
59880	&	41.4	&	19.8	&		$1.60^{+0.14}_{-0.13}$	&	$2.61	^{+	1.45	}_{	-1.07	}$	&	CXO/ACIS-S	&	Matthews$^2$ \\
59887.6	&	47.5	&	18.8	&		$1.18^{+0.10}_{-0.13}$	&	$0.94	^{+	0.58	}_{	-0.43	}$	&	CXO/ACIS-S	&	Matthews$^2$ \\
59889.2	&	48.7	&	4.4	&		$0.97^{+0.51}_{-0.39}$	&	$2.75	^{+	0.19	}_{	-0.18	}$	&	Swift/XRT	&	Schulze$^1$ \\
59900.5	&	57.7	&	2	&		$1.91^{+1.05}_{-0.78}$	&	$1.96	^{+	0.16	}_{	-0.15	}$	&	Swift/XRT	&	Schulze$^1$ \\
59905.4	&	61.6	&	18.1	&	$	1.41^{+0.12}_{-0.13}$	&	$1.45	^{+	0.14	}_{	-0.13	}$	&	CXO/ACIS-S	&	Matthews$^2$ \\
59914.2	&	68.7	&	5.8	&		$0.69^{+0.42}_{-0.31}$	&	$1.72	^{+	0.16	}_{	-0.15	}$	&	Swift/XRT	&	Schulze$^1$ \\
59941.1	&	90	&	37.8	&		$0.39^{+0.05}_{-0.04}$	&	$0.47	^{+	0.06	}_{	-0.05	}$	&	CXO/ACIS-S	&	Ho$^3$ \\
59974.9	&	117	&	39.6	&		$0.17^{+0.03}_{-0.03}$	&	$0.20	^{+	0.04	}_{	-0.03	}$	&	CXO/ACIS-S	&	Matthews$^2$\\
\enddata
\end{deluxetable}
\footnotesize
\vspace{-1cm}
 \hspace{1cm}$^1$ Swift Obs IDs 00015367001 - 00015367014 \\
 
 \hspace{1cm}$^{2}$ CXO Obs IDs 26641 - 26645 \\

 \hspace{1cm}$^{3}$ Joint fit of CXO Obs IDs 27639 \& 27643 \\
\end{document}